\documentclass{aa}
\usepackage{graphicx}

\begin{document}

\title{GMRT Observations of the Group Holmberg 124: Evolution by Tidal Forces and Ram Pressure ?}
\author{N. G. Kantharia,  S. Ananthakrishnan,  R. Nityananda,  Ananda Hota\thanks{Also Joint
Astronomy Programme, Dept. of Physics, Indian Institute of Science, Bangalore - 560012, India}}
\titlerunning{Tidal Force and Ram Pressure in Ho 124}
\authorrunning{Kantharia et al.}  
\institute{National Centre for Radio Astrophysics, 
Tata Institute of Fundamental Research, \\
Post Bag 3, Ganeshkhind,  Pune - 411007, India}
 
\offprints{N.G.Kantharia \email{ngk@ncra.tifr.res.in}}

\date{Received 27 October 2004 / Accepted 21 January 2005}
\abstract{We report radio continuum and 21cm HI observations using the Giant Metrewave Radio
Telescope (GMRT) of the 
group Holmberg 124 (Ho 124) comprising four late-type galaxies, 
namely NGC 2820, Mrk 108, NGC 2814 and NGC 2805.  The three galaxies, NGC 2820, Mrk 108
and NGC 2814 which are closely located in the sky plane have clearly undergone
tidal interactions as seen from the various morphological tidal signatures and debris. 
Moreover we note various features in the group members which we believe might be due to 
ram pressure. \\ 
In this paper, we describe four interesting results 
emerging from our observations: {\bf a)} detection
of the tidal radio continuum bridge at 330 MHz connecting the galaxies NGC 2820+Mrk 108 with NGC 2814.  
The radio bridge was discovered at 1465 MHz  by van der Hulst \& Hummel (\cite{hulst}).
We find that the bridge has a fairly steep spectrum with a spectral index $\alpha$ 
($S \propto \nu^{\alpha}$) of $-1.8^{+0.3}_{-0.2}$ which is
much steeper than the $-0.8$ quoted by van der Hulst \& Hummel (\cite{hulst}).
{\bf b)} Detection of other tidal features like the tilted HI and radio continuum 
disk of NGC 2814, a HI streamer and a radio continuum tail arising from the south of NGC 2814.
We also report the detection of a possible tidal dwarf galaxy in HI.  
{\bf c)}  Sharp truncation in the HI distribution in the south of NGC 2820 and in the HI \&
radio continuum distribution in the north of NGC 2814.  The optical disks in both the
cases look undisturbed.
As pointed out by Davis et al (\cite{davis}), ram pressure affects
different components of the ISM in varying degrees.
Simple estimates of pressure in different components of the interstellar medium 
(radio continuum, H$\alpha$ and HI) in NGC 2820 indicate that ram pressure will significantly
influence HI.
{\bf d)} Detection of a large one-sided HI loop to the north of NGC 2820.  No radio continuum
emission or H$\alpha$ emission is associated with the HI loop.  We discuss various scenarios
for the origin of this loop including a central starburst, ram pressure
stripping and tidal interaction.  
We do not support the central starburst scenario since it is
not detected in ionized gas.  Using the upper
limit on X-ray luminosity of Ho 124 (Mulchaey et al. \cite{mulchaey}), we estimate
an upper limit on the IGrM density of $8.8\times10^{-4}$ cm$^{-3}$. 
For half this electron density, we estimate the ram pressure force of 
the IGrM to be comparable to the gravitational pull of the disk of NGC 2820.
Since tidal interaction has obviously
influenced the group, we suggest that the loop could have formed by ram pressure 
stripping if tidal effects had reduced the surface density of HI in NGC 2820.  

From the complex observational picture of Ho 124 and the numerical estimates, we suggest 
that the evolution of the Ho 124 group may be governed by both
tidal forces due to the interaction and the ram pressure due to motion of
the member galaxies in the IGrM and that the IGrM densities should not be too low
(i.e. $\ge 4\times10^{-4}$).  However this needs to be verified by other observations.

 }

\maketitle

\section{Introduction}

In an ongoing project of studying the radio emission from disk galaxies 
using the GMRT, the poor group of galaxies known as Ho 124 has been observed.  
Ho 124 consists of four galaxies: an inclined SBc galaxy,
\object{NGC 2820} (\object{UGC 4961}), a Markarian galaxy, \object{Mrk 108} 
(\object{NGC 2820a}), 
an IO galaxy, \object{NGC 2814} and an almost face-on Sc galaxy, \object{NGC 2805} (\object{UGC 4936}).  
Since the first three galaxies lie within a few arc minutes of each other, we refer to them as
the triplet in the paper.  NGC 2805 lies about $8'$ to the south-west of the triplet.

This group is an interesting multiple interacting system. 
It was the first system in which a radio continuum bridge was detected
(van der Hulst \& Hummel \cite{hulst}). 
Although HI bridges and optical bridges had long before been
detected, no radio continuum emission had been detected leading to the belief 
that magnetic fields play little role in confining the bridges. 
Since then many other interacting galaxies have shown the presence of
radio bridges, e.g. the Taffy galaxies (Condon et al. \cite{condon}).
Bridges, tails and arcs have been detected from many other interacting systems.  Following
Toomre \& Toomre (\cite{toomre}), two long tails
are expected if two galaxies of comparable masses interact 
A bridge extending from one galaxy to other is expected if one galaxy is massive and the other
has a fraction of mass of the massive partner 

Many of the interacting systems occur in groups of a few galaxies, also known as poor groups.
Moreover, although the gravitational
interaction and ram pressure stripping of gas in members of clusters has been fairly
well-studied, less is known about these processes in groups.  The IGrM densities
are at least an order of magnitude lower than the intracluster medium (ICM). 
Hence, processes like ram pressure stripping and galaxy harrassment which play an important
role in the cluster evolution are not expected to be important in groups
(Mulchaey \cite{mulchaey4}). 
The first X-ray detection of the IGrM was made only a decade ago by Mulchaey et al. 
(\cite{mulchaey1}).
A more extensive X-ray survey of groups using ROSAT data was carried 
out by Mulchaey et al. (\cite{mulchaey2}) from 
which emerged the result that groups with at least one early-type galaxy have higher
X-ray luminosities than groups with only late-type galaxies.  Mulchaey et al. (\cite{mulchaey2})
gave some possible reasons including that the IGrM of groups with 
only late-type members had either lower temperatures
or lower densities.  In this paper, we present radio continuum observations 
at 240, 325, 610 and 1280 MHz and HI 21 cm observations using GMRT of one such group 
Ho 124 which consists of only late-type galaxies.  No X-ray emission was detected
from this group by Mulchaey et al. (\cite{mulchaey2}, \cite{mulchaey}).
We show that the IGrM densities in the group Ho 124 consistent with this upper limit
could still be sufficient to determine the evolution of the members via ram pressure stripping.

Additionally, we report the detection of a tidal bridge connecting the triplet
in radio continuum at 325 MHz and marginal detection at 240 MHz and 610 MHz.  
We also detect HI 21 cm emission from the bridge and a large
one-sided HI loop to the north of NGC 2820.  Bosma et al. (\cite{bosma}) have studied this
group in radio continuum, 21cm HI and in the optical band whereas
Artamonov et al. (\cite{artamonov}) have studied the group using UBV photometry.  
Optical properties of the group members can be found in Table 1 of Bosma et al. (\cite{bosma}) 
and in Table 1 of Artamonov et al. (\cite{artamonov}).

The plan of the paper is as follows.  In section 2, we discuss the observations, data analysis
and results.  In section 3, we discuss the various morphological features in
the group which we believe are due to the tidal interaction and in section 4, 
discuss various possible scenarios for
the origin of the HI loop in NGC 2820.  In section 5, we present a discussion
of our results and in section 6 we present a summary. 

Bosma et al. (\cite{bosma}) have used a distance of 24 Mpc to the group based on the mean
heliocentric radial velocity of 1670 km\,s$^{-1}$ and a Hubble constant of 75 km\,s$^{-1}$ Mpc$^{-1}$.
At this distance, $1'$ corresponds to 7 kpc.
We use the Bosma et al. (\cite{bosma}) values in the paper.  

\section{Observations, Data analysis and Results}

\begin{table*}
\caption{Observation Details}
\begin{tabular}{lccccc}
\hline\hline
Parameter & 1280 MHz  & 610 MHz & 330 MHz & 240 MHz & HI \\
\hline
Date of observation & 16/7/2002 & 6/9/2002 & 19/8/2002 & 6/9/2002 & 28/10/2003\\
On-source telescope time &  4 hrs  & 2.5 hrs & 3.3 hrs & 2.5 hrs & 5 hrs \\
Effective bandwidth & 9.3 MHz & 4 MHz  & 9.3 MHz & 4 MHz & 64 kHz \\
Phase calibrator & 0834+555 & 0834+555 & 0834+555 & 0834+555 & 0834+555 \\
Flux density of ph cal & $8.4\pm 0.13$ Jy & $8.05 \pm 0.15$ Jy  & $9.36\pm 0.25$ Jy & $9.1 \pm 0.3$ Jy
& $7.01 \pm 0.25 $ Jy  \\
Synthesized beam$^1$ &$19.9''\times 14.1''$ \& $6.5''\times 4.7''$  & $21.9''\times 13''$ & $19.9''\times 14.1''$ 
& $32.9''\times 14''$ & $16.2''\times 15.2''$  \\
PA &  $49^\circ.1$ \& $46^\circ.2$ & $69^\circ.9$ & $49^\circ.1$ & $82^\circ.1$ & $25^\circ.8$ \\
Continuum/line rms & 0.09 mJy/b \&  0.08 mJy/b & 0.4 mJy/beam & 1 mJy/beam & 1.9 mJy/beam & 0.2 mJy/beam\\ 
\hline
\end{tabular}
\label{tab1}

$^1$ this is the effective beamwidth of the images used in the paper and in most
cases is larger than the best achievable.
\end{table*}

\subsection{Radio Continuum}

\begin{figure*}
\centering
\resizebox{\hsize}{!}{\includegraphics{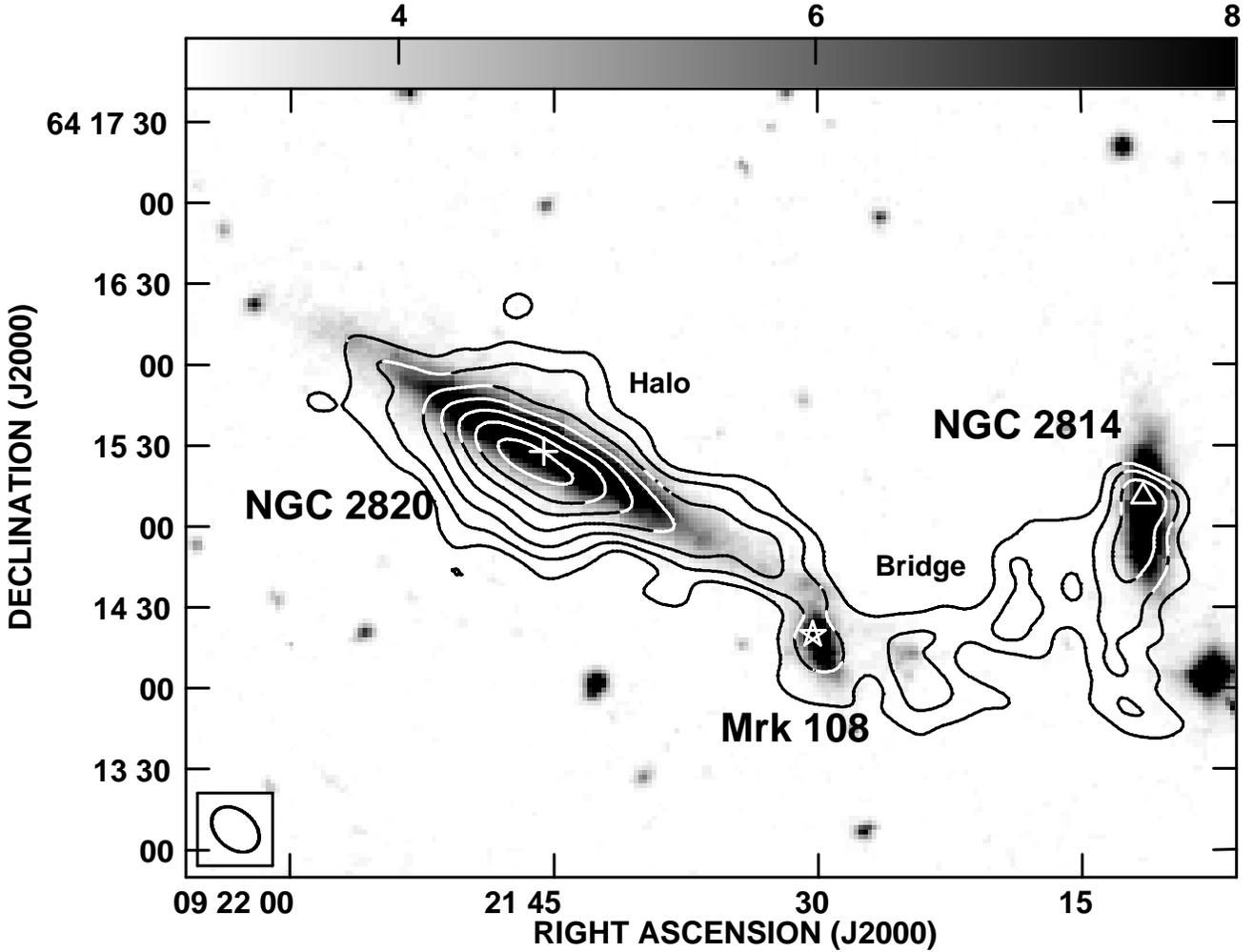}}
\caption{{\bf (a)} 330 MHz radio continuum contours superposed on the DSS grey scale image.
Note the bridge connecting the triplet and the halo emission.  The angular resolution of the
image is $19.9''\times14.2'', PA=49^\circ.1$ and rms noise is 1 mJy/beam.  The
first contour is plotted at 3 mJy/beam and then contours increase in steps of $\sqrt2$. 
The optical centres of the three galaxies are marked by a symbol.  }
\label{fig1}
\end{figure*}

The multi-frequency radio observations at 240, 330, 610 and 1280 MHz 
were conducted using the GMRT (Swarup et al. \cite{swarup}, 
Ananthakrishnan \& Rao \cite{ananth}) 
which consists of 30 antennas of 45m diametre each
scattered over a 25 km region. 
The observational details are listed in the first four columns of Table \ref{tab1}.  
These observations followed the sequence of interspersing 20 minutes on-source
scans by a 5 minutes scan on the phase calibrator.  The bandpass-cum-amplitude calibrator 
(3C 147) was observed in the beginning and at the end for half an hour each.  

The data was imported as a FITS file to NRAO AIPS software for further analysis.  
The general procedure followed at all bands included editing out
corrupted data, gain calibration of one spectral channel data, bandpass
calibration and channel averaging to obtain the continuum database.  These
were then imaged and CLEANed to obtain the final image.
Wide-field imaging was used 
at 610, 330 and 240 MHz.  We divided the primary beam into 9 facets for 610 MHz,
and into 25 facets at 330 MHz and 240 MHz.  The data were also 3-d self-calibrated. 
We used a uv taper of $12 k\lambda$
and a uv range of $15 k\lambda$ with robust weighting (ROBUST=0). 
Natural weighting did not seem to improve the image quality at these
three frequencies, but degraded the beam and hence was not used.
The 330 MHz and 240 MHz images are dynamic range limited.  We have obtained
a dynamic range of 2500 at 330 MHz.  At 1280 MHz,  a maximum uv
baseline of 60 $k\lambda$ was used with natural weighting.  We expect the flux density errors
at all frequencies to be less than 10 \%.  
All the images have been corrected for the gain of the primary beam.

The low resolution image at 330 MHz clearly showing the bridge
and the high resolution image at 1280 MHz showing fine structure in
the galaxies are shown in Fig \ref{fig1}.
The images at 240 and 610 MHz look fairly similar to the 330 MHz map
and hence are not presented.  We have used maps of similar resolution at
610 MHz and 330 MHz for estimating the spectral index. 

We have detected the triplet in radio continuum at all the observed frequencies.
Additionally, a bridge connecting the triplet 
is also detected at 330 MHz.  This bridge was first reported by van der Hulst
and Hummel (\cite{hulst}) at 1465 MHz.  
We have marginal detection of the bridge at 610 and 240 MHz.  
We have verified that although some short spacings are missing at 1280 MHz, this does not
completely resolve out the bridge.  However, alongwith the low brightness
sensitivity, this made it difficult to detect the bridge with the present data.
Faint radio emission is detected at 330 MHz from NGC 2805 (see Fig \ref{fig5} (b)).
This emission bears little resemblance to the optical emission (see Fig 5 (b)).

\addtocounter{figure}{-1}
\begin{figure}
\centering
\caption{
{\bf (b)} High resolution image of the group at 1280 MHz in contours is
superposed on the H$\alpha$ emission represented by the grey scale. 
The angular resolution of the radio image
is $6.54''\times 4.70''$, PA $= 46.21^{\circ}$ and rms noise is $80\mu$Jy.
The first contour is at 0.27 mJy/beam and then contours increase in steps of
$\sqrt2$.  The H$\alpha$ image is from Gil de Paz et al. (\cite{gil}).}
\label{fig1}
\end{figure}

The radio centre of NGC 2820, an almost edge-on galaxy
with inclination $\sim 84^{\circ}$ (Hummel \& van der Hulst \cite{hummel}), 
coincides with the optical centre within $5''$ (see Fig \ref{fig1} (a)).  
The flux densities of the galaxies at different frequencies and the galaxy-integrated
spectral index between 330 and 610 MHz are listed in Table \ref{tab2}.  

\begin{figure*}
\centering
\caption{ HI emission detected at different velocities is shown
in the panels.  The beam is plotted in the bottom left corner of
the first panel.  The first contour is at $2.4$ mJy/beam and then
it increases in steps of $\sqrt2$.  The grey scale is from 0.1 to 15 mJy/beam. 
Note that the HI in NGC 2814 is moving at velocities between 1660 to 1767 km\,s$^{-1}$
whereas the HI gas in Mrk 108 is moving at velocities between 1417 and 1444 km\,s$^{-1}$.
HI is detected from the bridge.  The optical positions
of NGC 2820, Mrk 108 and NGC 2814 are shown by a cross, star and triangle respectively. 
The cube has been smoothed in the velocity axis and every alternate channel is plotted here.}
\label{fig2}
\end{figure*}

\begin{table}
\caption{Radio flux densities of the triplet }
\begin{tabular}{clcccc}
\hline\hline
& Galaxy & \multicolumn{3}{c}{S mJy at $\nu$ MHz} & $\alpha^{330}_{610}$ \\
&  & 1280$^1$ & 610 & 330   &  \\
\hline
1 &NGC 2820+  & 35  & $116$  & $227$  &   $-1.06$  \\
& Mrk 108 &  & (4) & (4) &  \\
2& NGC2820 & & 19.9 & 27 &  $-0.5$  \\
 & peak           & & (0.7) & (1) &  \\
3 & NGC 2814  & 6.7   & $19$   & $42$   &   $-1.25$ \\
          &   & (1.6) & $(4)$  & (6) &  \\
4 & NGC 2814  & & 5.2 & 7.5 &  $-0.6$  \\
 & peak            & & (0.7) & (1) &   \\
\hline
\end{tabular}
\label{tab2}

$^1$ the flux density of NGC 2820 at 1280 MHz that we find is lower than the value quoted by others.  
Hummel \& van der Hulst (\cite{hummel}) estimate a flux density of $60\pm5$ mJy at 1.465 GHz
with the VLA whereas Bosma et al. (\cite{bosma}) have estimated a flux density of $48\pm5$ mJy
at 21cm using the WSRT.  Condon et al. (\cite{condon90}) find a flux
density of 64.2 mJy at 1.49 GHz.  Since the flux density, we estimate, is lower 
we do not quote the image errors which are comparatively insignificant. \\
\end{table}

Halo emission is detected around NGC 2820 at all the observed frequencies and is prominent at
330 MHz (see Fig \ref{fig1} (a)) which is our most sensitive low frequency.  
In their study of radio emission from six edge-on galaxies, Hummel \& van der Hulst (\cite{hummel}),
found that NGC 2820 had a radio halo with the largest z-extent (10\% peak
height of about 3.4 kpc) which they attributed
to gravitational interaction with its companions.  
We find the 10\% peak flux level z-extent of the radio halo is 4.2 kpc.  
The half power thickness of the halo is 2.2 kpc which is double the typical value.
We estimated the spectral index between 330 and 610 MHz (from similar angular resolution
maps) at a few positions in the halo and find it to be $-1.5$.  

NGC 2814 shows halo emission which is tilted with respect to the
stellar disk traced by the DSS optical image (see Fig \ref{fig1}(a)).  We find 
that the global spectral index of the 
galaxy is $-1.25$ whereas that of the radio peak is $-0.6$ (see Table \ref{tab2}).  
It is difficult to separate the halo emission from the disk emission for this
relatively small galaxy.

The radio power of NGC 2820 is $1.2\times10^{22}$ Watt-Hz$^{-1}$ 
and that of NGC 2814 is $2.6\times 10^{21}$ Watt-Hz$^{-1}$ (estimated at 330 MHz).
We also estimated the q factor which gives the ratio of FIR flux density to the radio continuum
flux density at 1.4 GHz for NGC 2820 following Helou et al. (\cite{helou}).
We find that q $= 2.02$ for NGC 2820.  This value is consistent
with the quoted value for spiral galaxies of q $= 2.3$ 
with a rms scatter of 0.2 (Condon \cite{condon1}).  Thus,
NGC 2820 follows the FIR-radio correlation.

\subsection{21 cm HI}

The details of the 21cm HI observations  are listed in Table \ref{tab1}.
These data were initially analysed in a way similar to the continuum data 
sans self calibration.
The continuum emission from the line data was removed and spectral
channels imaged to generate a cube.
We used a uv taper of $12 k\lambda$ and uv range of $15 k\lambda$
with robust weighting (ROBUST = 0) to obtain the final cube. 
The beamwidth is $16.2''\times 15.2''$ with a PA$=25.7^{\circ}$
which greatly improves on the arcmin resolution of Bosma et al. (\cite{bosma}).

\begin{figure}
\centering
\resizebox{\hsize}{!}{\includegraphics{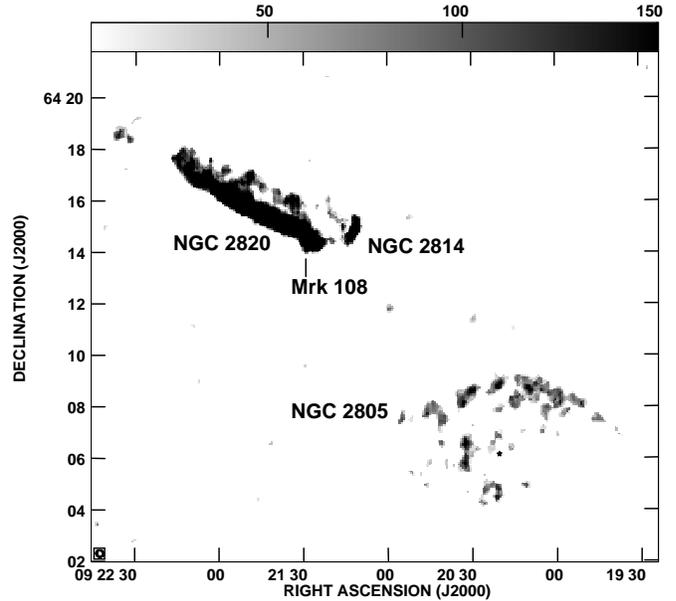}}
\caption{The HI moment zero map of the group Holmberg 124.  The triplet comprising 
of NGC 2820, Mrk 108
and NGC 2814 lies in the north-east whereas NGC 2805 is the face-on member seen in the south-west.
Note that NGC 2805 was close to the half power point of the  primary beam at 21 cm.} 
\label{fig3}
\end{figure}

\begin{figure*}
\centering
\resizebox{\hsize}{!}{\includegraphics{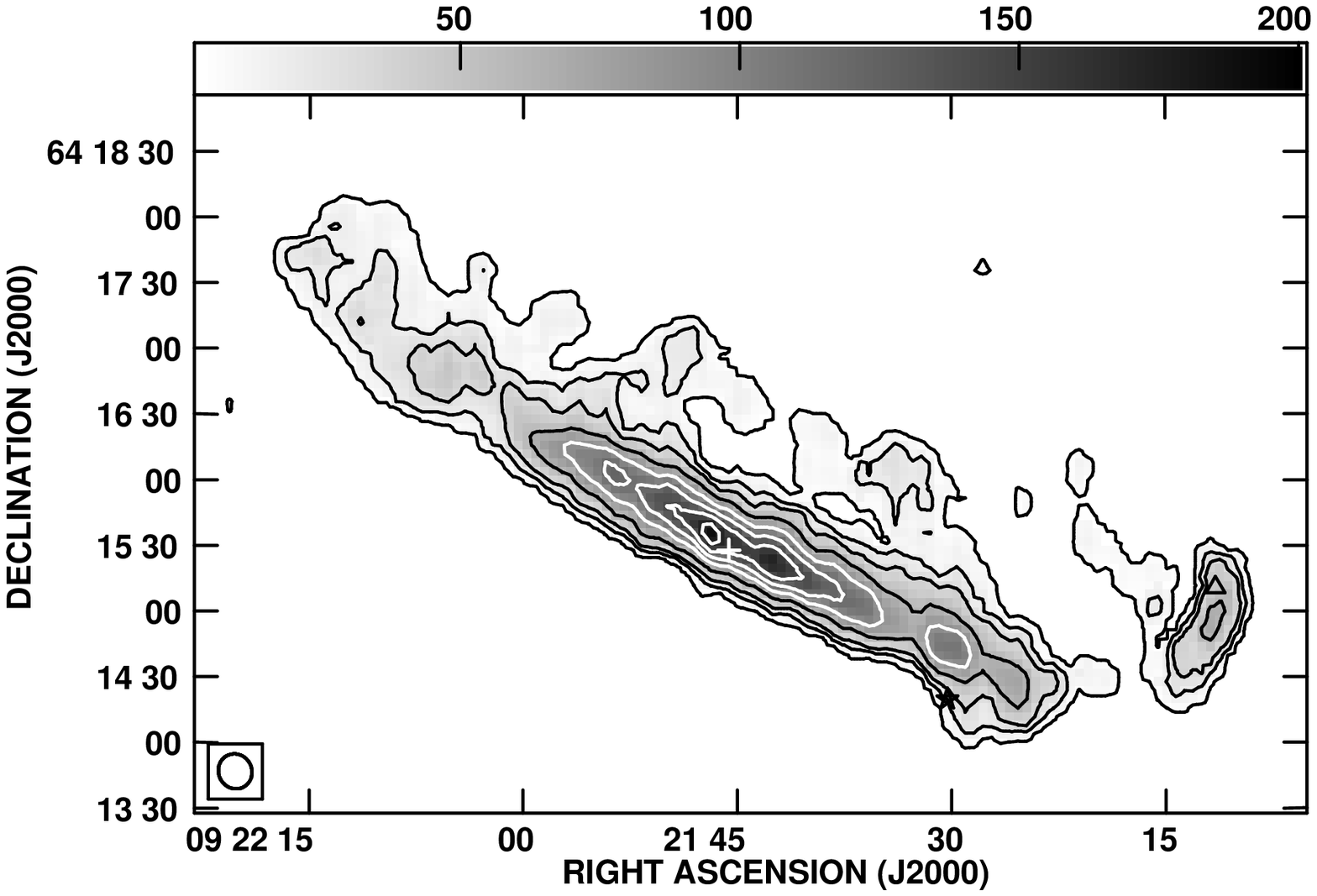}}
\caption{{\bf (a)} Column density map of the triplet obtained from the
moment zero map.  The lowest contour
is $0.44\times10^{20}$ cm$^{-2}$.  The next two contours
are at an inverval of $1.4\times10^{20}$ cm$^{-2}$  and then
the contours increase in steps of $2.8\times10^{20}$ cm$^{-2}$.
The grey scale ranges from $0.4\times10^{20}$ cm$^{-2}$ to $20\times10^{20}$ cm$^{-2}$}.
\label{fig4}
\end{figure*}

\addtocounter{figure}{-1}
\begin{figure}
\centering
\resizebox{\hsize}{!}{\includegraphics{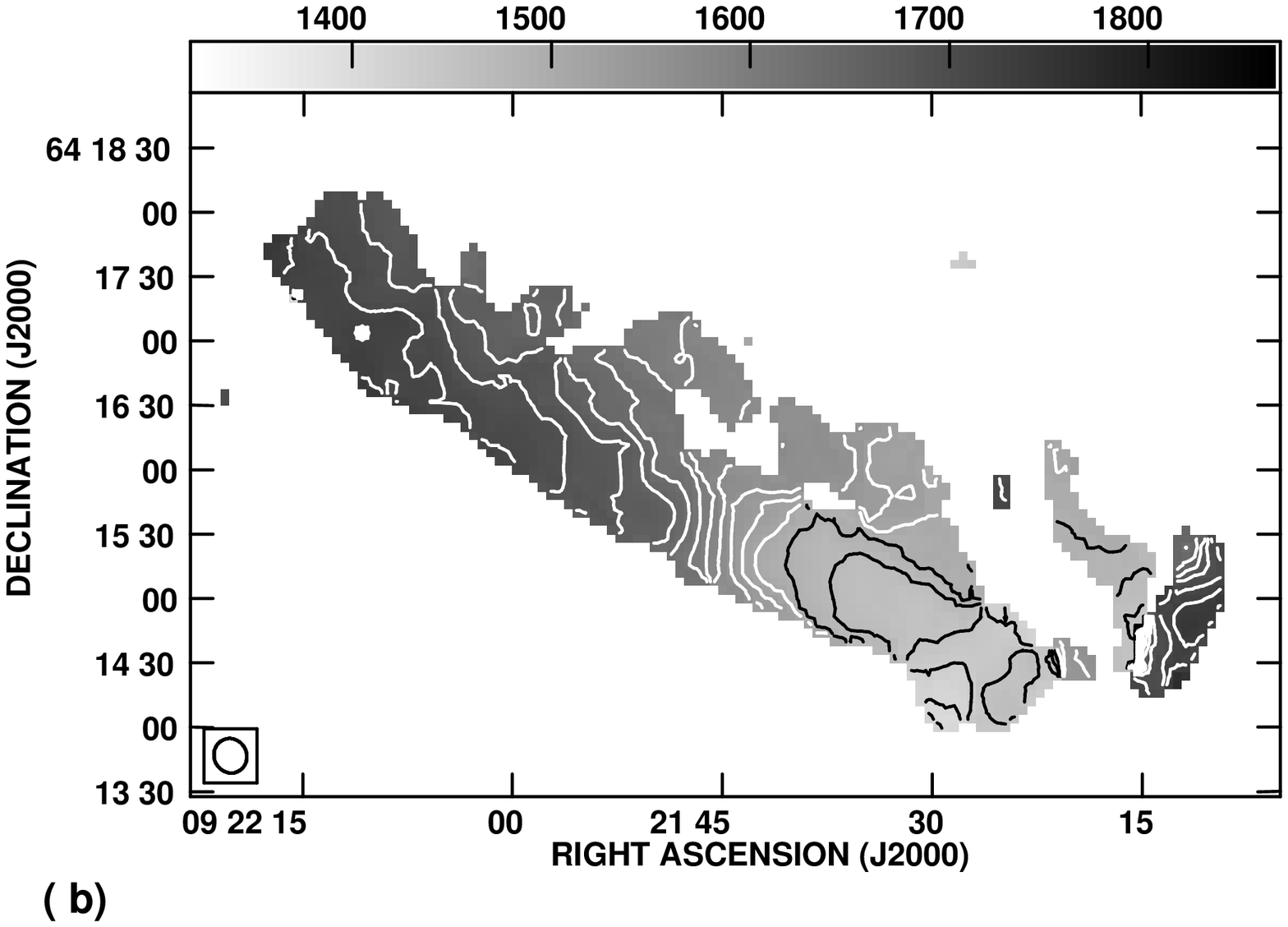}}
\resizebox{\hsize}{!}{\includegraphics{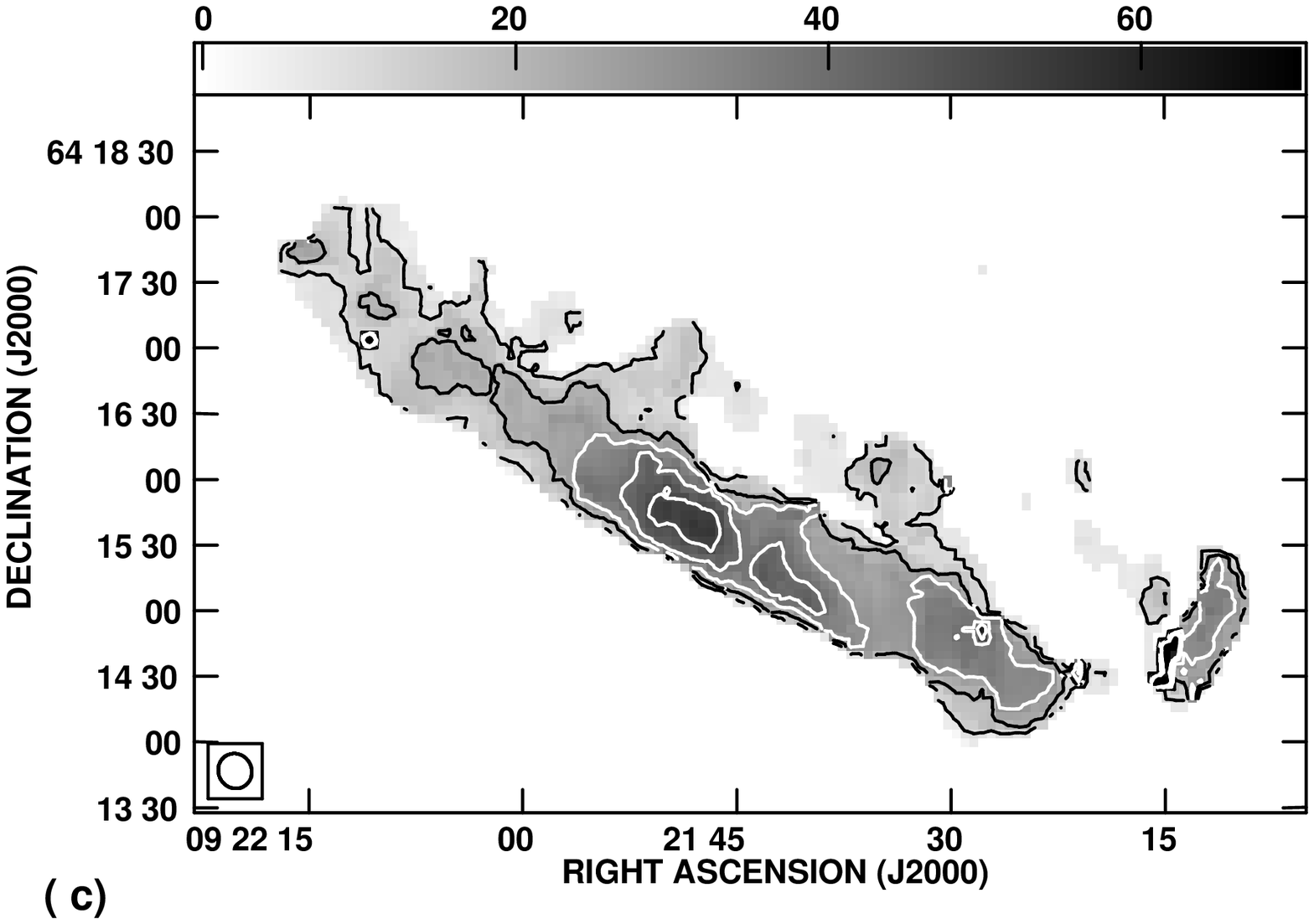}}
\caption{{\bf (b)} (top panel) First moment map of the HI emission showing the
velocity field in the triplet.  The grey scale ranges from 1323 km\,s$^{-1}$ to
1862 km\,s$^{-1}$.  The contours are plotted from 1410 km\,s$^{-1}$ (black contour)
to 1730 km\,s$^{-1}$ in steps of 20 km\,s$^{-1}$.  Note the distinct velocity field of Mrk 108.
{\bf (c)} Second moment map of the HI emission showing the line widths in the triplet.
The grey scale ranges from 0 to 70 km\,s$^{-1}$.  The contours are plotted from
10 to 50 km\,s$^{-1}$ in steps of 10 km\,s$^{-1}$.}
\label{fig4}
\end{figure}

We detected HI from all members of the group. 
The channel maps showing HI line emission detected at different velocities for the
triplet are shown in Fig \ref{fig2}.  
The column density map of the group (estimated assuming HI is optically
thin along the line of sight) is shown in Fig \ref{fig3}.  
A zoomed-in HI column density map for the triplet 
is shown in Fig \ref{fig4} (a).  
The first and second moment maps of HI
for the line emission are shown in Figs \ref{fig4} (b) and (c).
\begin{figure}
\centering
\caption{{\bf (a)} (top panel) Contours of the zero moment map of HI emission 
superposed on the grey scale image of DSS blue band image of NGC 2805.  Note the ridge of
HI emission in the north which coincides with a disturbed
spiral arm seen in the optical.  Note that HI is confined to the optical disk 
of the galaxy.  Since this galaxy lies close to the
half power point of the GMRT primary beam, we do not trust the HI column densities.
However the morphology except for the extended disk emission roughly follows what
Bosma et al. \cite{bosma} had reported.
{\bf (b)} Contours of 330 MHz radio emission superposed on the
grey scale DSS blue band image of NGC 2805.  Note the ridge of
star formation and sharp cutoff in the south-west visible in the DSS image.
The star marks the optical centre of the galaxy.  }
\label{fig5}
\end{figure}

The channel maps (see Fig \ref{fig2}) clearly show the rotation 
in the disk of NGC 2820.   
The HI disk extends more to the northeast than in the southwest. 
The velocity of the gas in NGC 2820 varies from $\sim 1710$ km\,s$^{-1}$ in
the north-east to $\sim 1445$ km\,s$^{-1}$ in the southwest (see Fig \ref{fig4} (b)).
Since the systemic velocity of the galaxy is 1577 km\,s$^{-1}$, this gives a
difference velocity of 133 km\,s$^{-1}$ in the northeast and 134 km\,s$^{-1}$ in the southwest.
The rotation speed of the gas is fairly symmetric over the centre unlike what
Bosma et al. (\cite{bosma}) found.  This is probably because 
the velocity of the HI gas in Mrk 108, which
we can clearly distinguish in our maps (see Fig \ref{fig2} - last two panels) 
due to our higher angular resolution is around
1410 km\,s$^{-1}$ and was presumably included within the Bosma et al. (\cite{bosma}) beam.  
It is of interest to note the
nature of the isovelocity contours.  In Fig \ref{fig4} (b), it is seen that the
contours in the northeast are different from those in the southwest
and appear to be kinematically disturbed. 
It appears that the HI gas in the northeast has been affected by the
interaction more than in the southwest.
A ring of HI surrounds the optical center (see Fig \ref{fig4} (a)).
HI condensation is also seen near Mrk 108 and it is likely associated with a star forming 
region seen in the H$\alpha$ image (see Fig \ref{fig7} (c)), possibly triggered by the interaction.
 
HI is detected from the bridge except for a small region.  
The mean HI column density in the bridge is $\le 4.4 \times 10^{19}$ atoms cm$^{-2}$.

HI emission shows interesting extraplanar features
in NGC 2820. A symmetric HI loop 
is observed to the north of the galaxy - the channel maps (Fig \ref{fig2}) 
between 1633 km\,s$^{-1}$ and 1525 km\,s$^{-1}$ clearly show the presence of extraplanar features.
The loop opens at the top giving it an appearance of
an outflow and then seems to turn back as if
the HI gas is falling back towards the disk.  The loop has enormous dimensions
with a width parallel to the galactic disk of about 17.5 kpc ($\sim2.5'$) and a height of about
4.9 kpc ($\sim 0.7'$).  
No detectable radio continuum emission is associated with the HI loop.
Moreover there are a couple of other protrusions visible to the east of the large HI loop.
Interestingly, no HI filaments or protrusions are observed arising from the 
southern side of NGC 2820 and the HI shows a smooth boundary.
Similarly, we find that the HI disk and radio continuum disk
of NGC 2814 are sharply truncated in the north of the galaxy whereas the optical emission
extends beyond.
The HI disk of NGC 2814, like the radio continuum disk, is inclined to the optical disk. 
A high velocity streamer is seen emerging, almost perpendicularly, from the south of 
NGC 2814.  This streamer is fairly long, extending to the northeast.
The heliocentric velocity of NGC 2814 is 1707 kms$^{-1}$ whereas the streamer has a line
of sight velocity range of $\sim 1452$ kms$^{-1}$ to $1510$ kms$^{-1}$. 
The velocity difference between the
streamer and NGC 2814 is more than 200 kms$^{-1}$.  The streamer velocity 
is more in tune with the velocity field seen in the southern parts of NGC 2820.

\begin{table*}
\caption{Galaxy parameters from HI data.  
A distance of 24 Mpc to the group has been used.
The systemic velocities of Mrk 108, NGC 2814, NGC 2805, the streamer and the HI blobs have been determined
by fitting a gaussian to the source-integrated HI profile. }
\begin{tabular}{lccccccc}
Parametre & NGC 2820  & Mrk 108 & NGC 2814  & NGC 2805 & streamer & HI 'blobs'& HI Loop\\
\hline\hline
Heliocentric velocity ($km\,s^{-1}$) &  1577 & 1417  & 1707 & 1745 &  1493  & 1725 & 1566  \\ 
Half-power width ($km\,s^{-1}$) & 350 & 52 & 136 & 99 & 72  & 42 & 137\\
Rotation velocity ($km\,s^{-1}$) &  175  & - & - &  &- & - &- \\
Inclination ($^{\circ}$) &  74 & - & - & 20$^1$ & - & - & - \\
Position angle ($^{\circ}$) &  66 & - & - & & - & - & - \\
Dynamical centre (J2000) & 09h21m45.6s & - & - & & - & - & -\\
                         & 64d15m31s   & - & - & & - & - & - \\
Linear size ($kpc$) & 47.6 & 3 & 10.4 & 60 & 12.7 & 3.5 & 17.5 kpc \\
HI mass $M_{HI}$ ($10^9 M_\odot$) & 6.6 & 0.061 & 0.34 & $5.3^2$ & 0.13 & 0.11 & 0.6\\
Dynamical mass $M_{dyn}$ ($10^9 M_\odot$) & 170 & 0.94 & 22 & 584 & - & 1.4 & -\\
$M_{HI}/M_{dyn}$ (\%) & 3.9  & 6.5  & 1.5 & - & - & 7.9 & - \\
\hline
\end{tabular}

$^1$ From Bosma et al. (\cite{bosma}). This value is used in estimating the HI mass and dynamical mass \\ 
$^2$This estimate is much less than 
the value of $12\times10^9$ $M_\odot$ of Bosma et al. (\cite{bosma}) and is likely because
the galaxy is close to the half power points of the GMRT primary beam. \\ 
\label{tab3}
\end{table*}

The HI distribution of NGC 2805 (see Fig \ref{fig5} (a)) is asymmetric 
with larger column densities and higher radial velocities in the northern regions
as compared to the southern parts.
Since this galaxy was close to the half power point 
of the GMRT primary beam in our 21cm HI image, we cross checked the observed morphology 
with the Bosma et al. (\cite{bosma}) images. 
We find the two maps correlate well and the paucity of HI in the southern
parts is real and not an artifact of the primary beam cutoff.  However we are 
insensitive to the large scale HI emission seen from the face-on disk by Bosma et al. (\cite{bosma}). 

Global HI line profiles were obtained for all the galaxies and HI features.  A gaussian function
was fitted to the observed profiles (except for NGC 2820 which shows
a classical double-humped HI profile with a sharp fall-off) and the resultant
parameters were used to derive various physical quantities (Table \ref{tab3}).
The systemic velocity, rotation velocity, inclination and dynamical centre of NGC 2820
are results from running the task, GAL in AIPS on the velocity field of the
galaxy.  We model the observed data with a Brandt curve purely as a fit; it
reproduces the solid body rotation in the central regions fairly well.
We obtain a rotation velocity of 175 km\,s$^{-1}$ for NGC 2820.  The rotation curve
is shown in Fig \ref{fig8}.
The HI mass was calculated
from the velocity-integrated line strength whereas the dynamical mass was estimated using
$rv^2/G$.  We find that the rotation curve fit gives an inclination of $74^{\circ}$ and position
angle of $66^{\circ}$ for NGC 2820.  The optical heliocentric velocity of NGC 2820 is
1577 km\,s$^{-1}$ which is in good agreement with the value of $1574\pm10$ km\,s$^{-1}$ quoted
by Bosma et al. (\cite{bosma}).  
About 4\% of the total mass of NGC 2820 is in HI whereas
6.5\% mass of Mrk 108 appears to be in the form of HI (Table \ref{tab3}).  For
the HI blobs, we find that about 8\% of its dynamical mass is seen in the form of HI.
We find that NGC 2805 is massive with a total dynamical mass of $5.8\times10^{11}$ M$_\odot$.

The position-velocity (PV) curves plotted along and parallel
to the major axis of NGC 2820 are shown in Fig \ref{fig9}.
Fig \ref{fig9}(a) shows that the gas in the 
central 5.8 kpc of NGC 2820 exhibits solid body rotation.  
Some asymmetry is visible between 1600 and
1650 km\,s$^{-1}$.  The 
HI blob, NGC 2814 and the streamer  
are shown in the PV diagram (see Fig \ref{fig9} (b)) of a slice parallel to the major
axis of the galaxy.  The streamer is seen to be kinematically independent of
NGC 2814.

\section{Tidal Effects}
\label{tidal}
NGC 2820 appears to have had a retrograde interaction with NGC 2814 and the two galaxies
presently have a relative radial velocity of 130 km\,s$^{-1}$. 
Various morphological signatures which are likely due to the tidal interaction between 
NGC 2820, Mrk 108 and NGC 2814 are seen in our HI moment zero and radio continuum maps.  
The dominant signatures in HI are the streamer apparently emerging from NGC 2814 (but 
showing a different velocity field), 
the inclined disk of NGC 2814, the bridge between
NGC 2820 and NGC 2814 and the detection of HI blobs to the north-east of NGC 2820.
Star formation seems to have been triggered
in the disk of NGC 2820 close to Mrk 108, in Mrk 108 and in a small tail of 
NGC 2814  by the tidal interaction 
as can be identified on the H$\alpha$ image of the triplet (see Fig \ref{fig7} (c)).
Moreover, Artamonov et al (\cite{artamonov}) from their UBV photometric observations,
report enhanced star formation in Ho 124 due to tidal interaction.
The tidal features which are readily discernible in the 330 MHz map are
the steep spectrum radio bridge, the tilted radio disk of NGC 2814 and a 
radio tail issuing from NGC 2814 and extending southwards.  
However, the tidal origin of HI features like the
HI loop arising from the northern side of NGC 2820 and the small HI protrusions is not clear.
We discuss the origin of the loop in the next section.  

In this section, we briefly elaborate on some of the clear signatures of
tidal interaction discernible in our images. 

\subsection{Tidal Bridge }
van der Hulst \& Hummel (\cite{hulst}) were the first to detect a radio continuum bridge
connecting the triplet in Ho 124 at 1465 MHz.  Since then radio bridges 
have been detected in many other systems; a famous one being
the Taffy galaxies (Condon et al. \cite{condon}) in which one-half the 
total radio synchrotron emission of the system arises in the bridge.  

As shown in Fig \ref{fig1}(a), we have detected the bridge connecting the triplet at 330 MHz.  
Using the 1465 MHz result of van der Hulst and Hummel (\cite{hulst}) along with our
data at 330 MHz we estimate a spectral index of $-1.8^{+0.3}_{-0.2}$
for the bridge.  This spectral index is much steeper than the value of $-0.8$ quoted
by van der Hulst and Hummel (\cite{hulst}) which might possibly have been corrupted by
disk emission at their lower frequency.   
At 610 MHz and 240 MHz, we report marginal detection of the bridge
and the brightness of the bridge is consistent with the estimated spectral index.  

We estimated the size of the bridge from our 330 MHz image.
The projected linear extent of the bridge 
is 5.4 kpc.  Our estimate of the projected length of the
bridge is less than what van der Hulst \& Hummel (\cite{hulst}) estimated.
This is likely because we have not included the source west of
Mrk 108 which we believe is part of the disk of NGC 2820 and not the bridge. 
The width of the bridge in the sky plane is 2.1 kpc.
We assumed a similar extent for the bridge along the line-of-sight.  
Using equipartition and minimum energy arguments, 
we estimated a magnetic field of $\sim 3.4 \mu$G,
and a minimum energy density of $1.1 \times 10^{-12}$ ergs\,cm$^{-3}$.
The minimum energy in the bridge is $7.7\times 10^{53}$ ergs. 
The magnetic pressure of the bridge is about 2600 K\,cm$^{-3}$.
We have assumed that there is 100 times more energy
in protons than electrons for the above calculations.  A
few possible scenarios for confining the bridge were described in Ananthakrishnan et al. (\cite{ananth1}).  

We detected HI in the bridge except for a small region (see Fig \ref{fig4} (a)). 
This HI is moving at a line-of-sight velocity
of about 1545 km\,s$^{-1}$ which is different from the HI in the disk of NGC 2820
nearest to the extension of 1445 km\,s$^{-1}$ (see Fig \ref{fig4} (b)).  However the
velocity of HI in the bridge is closer to the systemic velocity of NGC 2820.  
If the gas in the bridge is moving with a velocity of 
100 km\,s$^{-1}$ (ie. $1545-1445$ km\,s$^{-1}$) with respect to the gas in NGC 2820, then
the kinematic age of the bridge is 46 million years.  If the gas is moving with a velocity larger
than 100 km\,s$^{-1}$, the kinematic age will be lower.  
The mean column density in the bridge is $\le 4.4 \times 10^{19}$ cm$^{-2}$.
Using a line-of-sight depth of 2.1 kpc for the
bridge, we find that the atomic density in the bridge is $< 0.006$ cm$^{-3}$.
If we assume a kinetic temperature of 5000 K, the thermal pressure of the
gas in the bridge would be only 30 K\,cm$^{-3}$ which is much less than the
magnetic pressure of the bridge. 
We obtain a result
similar to van der Hulst \& Hummel (\cite{hulst}) in that the magnetic field and relativistic
particles moving in it seem to dominate the bridge energetics.
It appears likely that the bridge is confined by an ordered magnetic field.  
No H$\alpha$ emission is detected from the bridge 
(Gil de Paz et al. \cite{gil}) (see Fig \ref{fig7} (c))
indicating that star formation has not been triggered in the bridge.  This is not surprising
since the column density of neutral atomic matter in the bridge is fairly low.  Little
molecular gas is therefore likely to be present in the bridge. 

\subsection{Tidal Effects on NGC 2814}
The disk of NGC 2814 has obviously been affected by tidal forces in its
encounter with NGC 2820.  
The optical disk is aligned almost north-south whereas the
radio continuum and HI disks are inclined towards the bridge clearly showing that
they have been affected by tidal forces (see Fig \ref{fig1}(a) and
Fig \ref{fig7}(a)) 
Also note the 'comma'-shaped HI disk and H$\alpha$ emission of NGC 2814.
The H$\alpha$ image of NGC 2814 (see Fig \ref{fig7} (c))
shows enhanced star formation in a small tail which is likely
triggered by the tidal interaction.  
Star formation, triggered by the tidal interaction is also observed in and close to Mrk 108.  
A tail is observed in the radio continuum issuing and extending to the south of NGC 2814
(see Fig \ref{fig1} (a)).  The spectral index of this tail is $\sim -1.6$ and the tail
is likely a result of the tidal interaction. 

\subsection{Tidal streamer }
A HI streamer is observed to arise from the southern end of NGC 2814 and extend
towards the north-east (see Fig \ref{fig4} (a)) but which is kinematically
distinct from the galaxy.
The different velocities (difference of about 200 kms$^{-1}$) 
support a projection effect.  A velocity gradient of about 5.3 km\,s$^{-1}$ kpc$^{-1}$
is observed along the streamer.  We note that the velocity field seen in
the streamer matches the velocities seen on the closer side of NGC 2820 and intriguingly
the shape of the tail matches the outer edge of NGC 2820.  
We believe that the streamer is HI gas which
has been stripped off NGC 2820 during the tidal interaction.  
Since we do not see an extra
radial velocity that the streamer might have picked up during the tidal encounter,
it might be in motion in the sky plane.  If we assume  that the streamer
was dislocated from NGC 2820 and picked up a velocity of $100$ km\,s$^{-1}$ 
due to the tidal interaction, then it would have taken
about 40 million years for the streamer to be at its current position.
The average column density in the streamer is
$4.4 \times 10^{19}$ cm$^{-2}$ and the mass is $\sim 9.1 \times 10^7 M_{\odot}$.  
The length of the streamer in the sky plane is about 12.6 kpc.  

\subsection{A tidal dwarf galaxy ?}
\begin{figure}
\centering
\resizebox{\hsize}{!}{\includegraphics{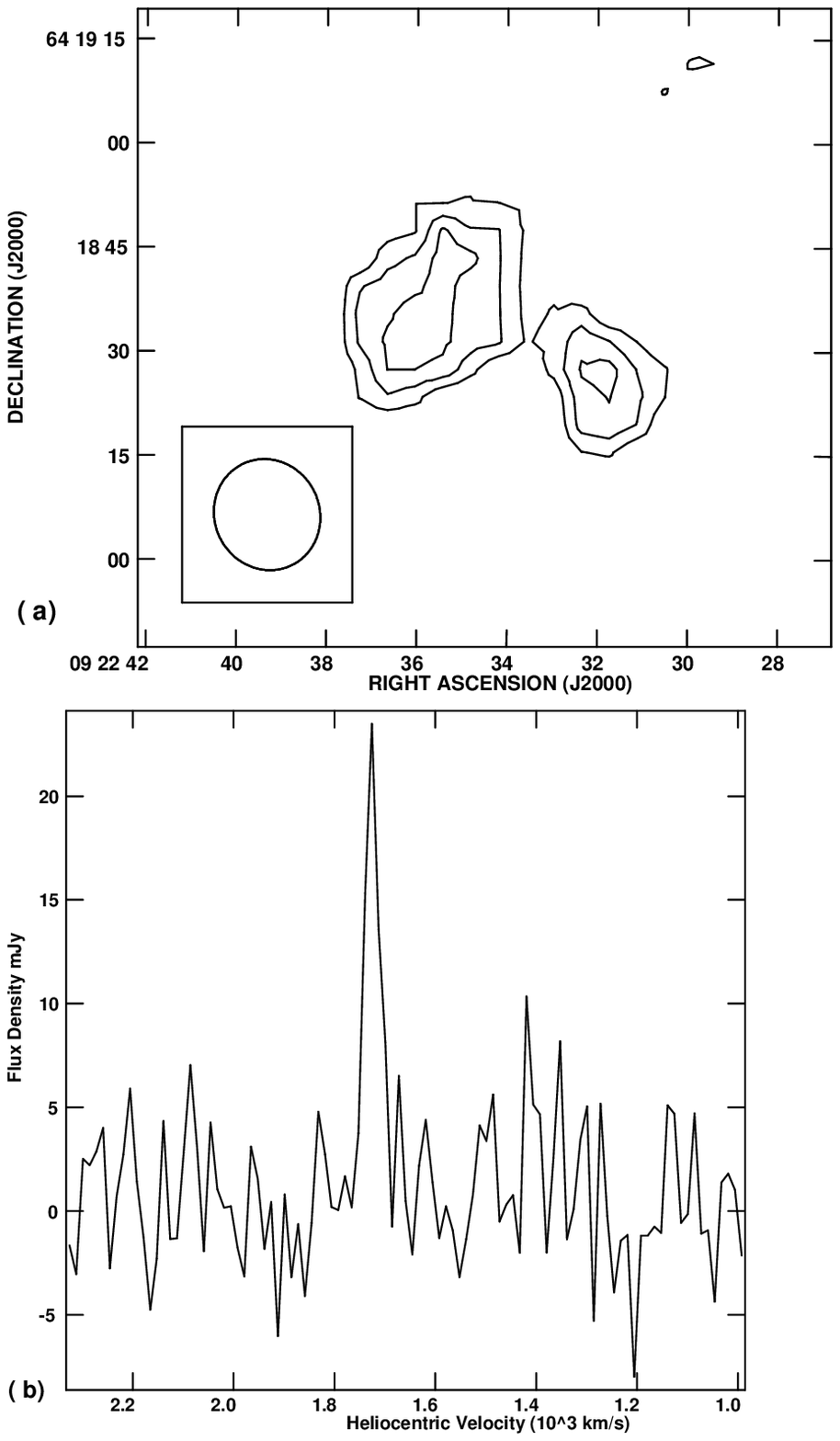}}
\caption{{\bf (a)} HI moment zero map of the HI 'blobs' which might be a tidal dwarf. 
The contours outline column densities of $4.4\times10{19}$ cm$^{-2}$,$8.8\times10^{19}$ cm$^{-2}$
nd $13.2\times10^{19}$ cm$^{-2}$.
{\bf (b)} The HI profile integrated over the HI 'blobs'.  }
\label{fig6}
\end{figure}

The HI blobs (see Fig \ref{fig6} (a)) 
detected to the north-east of NGC 2820 contain about $10^8 M_{\odot}$ of HI.
One possibility to explain these blobs is that it is a tidal dwarf galaxy. 
It would be interesting to obtain a deep H$\alpha$ image of this region
and check this possibility.
The spectrum integrated over the blobs is shown in Fig \ref{fig6}(b). 
No rotation is discernible in the
blobs.  We do not find any optical counterparts to the blobs in the DSS images. 
The HI velocity field of the blobs is a continuation of the velocity field
seen in the north-east tip of NGC 2820 probably indicating its origin.
The blobs are located about 11.5 kpc away from the north-east tip of NGC 2820.  

\section{The HI Loop}

\begin{figure}
\centering
\resizebox{\hsize}{!}{\includegraphics{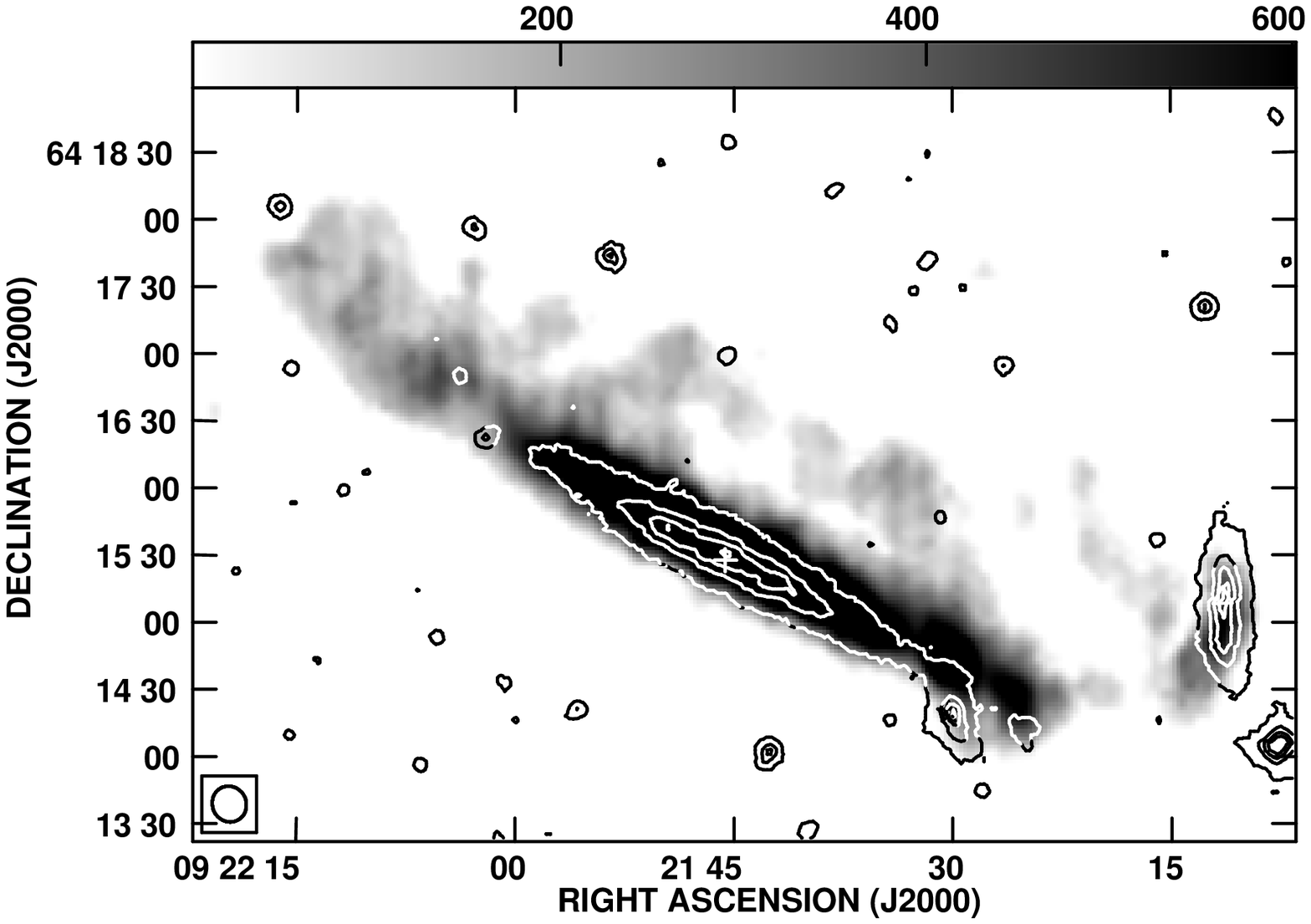}}
\resizebox{\hsize}{!}{\includegraphics{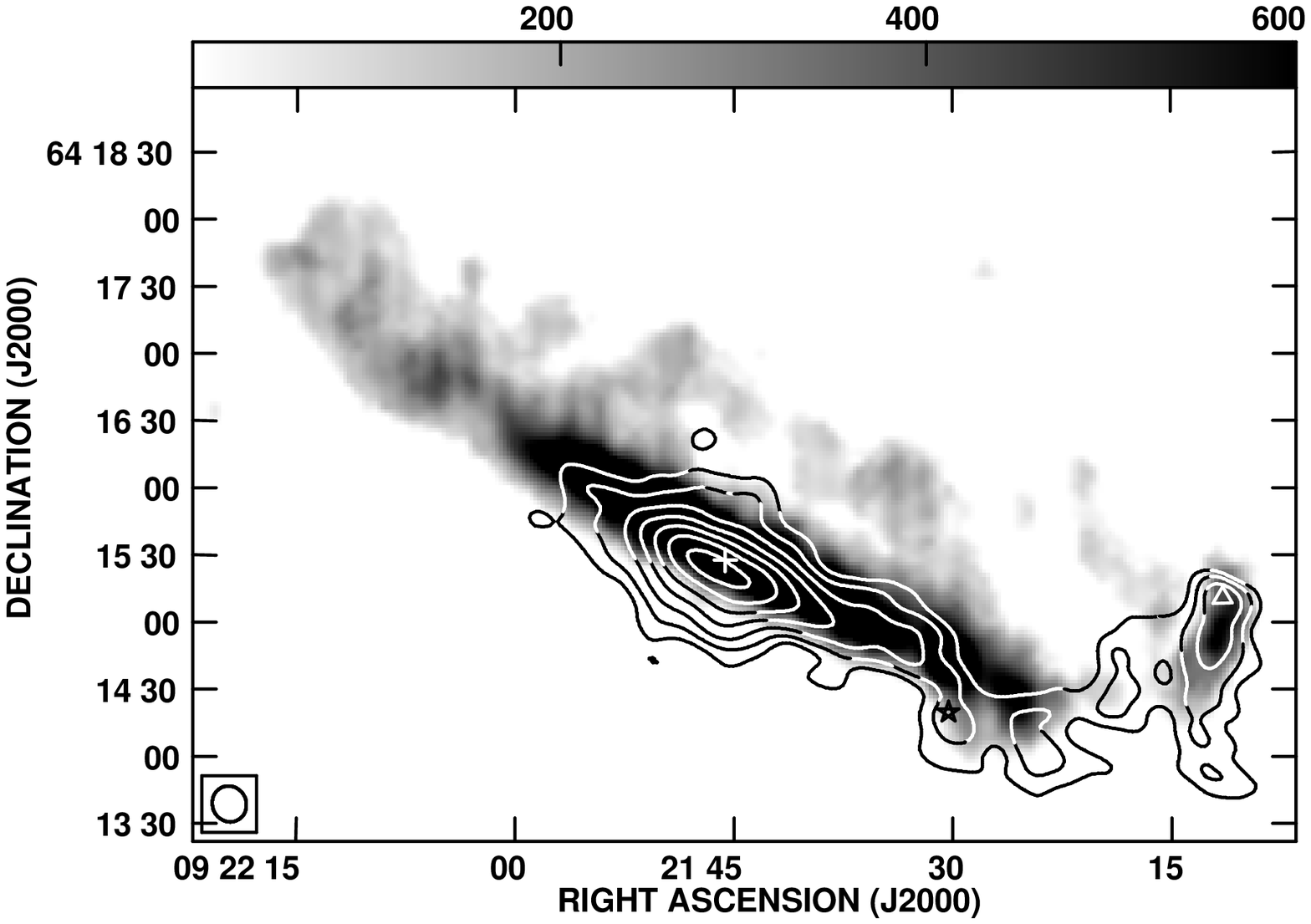}}
\caption{{\bf (a)} Grey scale of the HI column density superposed on the DSS
optical map (contours).  Note the extent of the HI compared to the optical continuum.
{\bf (b)} HI zero moment map (grey) superposed on the 330 MHz radio
continuum map. }
\label{fig7}
\end{figure}

\addtocounter{figure}{-1}
\begin{figure}
\centering
\caption{{\bf (c)} H$\alpha$ contours superposed on the moment zero map of HI in grey scale. 
Note the lack of H$\alpha$ emission in the bridge and in the loop.  The H$\alpha$ image is 
from Gil de Paz et al. (\cite{gil}).}
\label{fig7}
\end{figure}

\begin{figure}
\centering
\resizebox{\hsize}{!}{\includegraphics{fig8.eps}}
\caption{ Rotation curve of NGC 2820.  The solid line is 
the Brandt model fit to the data, results from which are noted in
Table \ref{tab3}.}
\label{fig8}
\end{figure}

\begin{figure*}
\centering
\resizebox{\hsize}{!}{\includegraphics{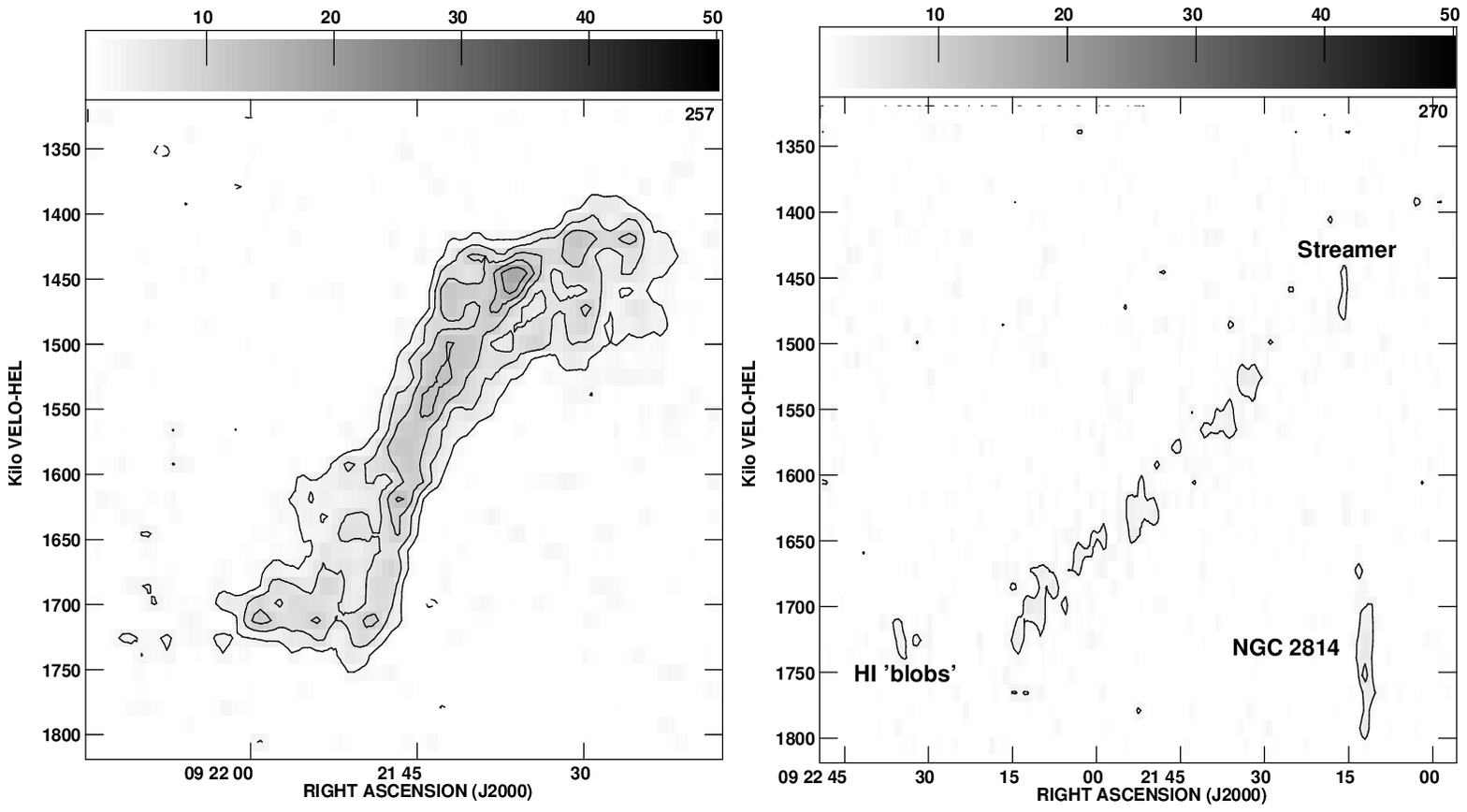}}
\caption{{\bf (a)} PV curve along the major axis of NGC 2820.  Solid body rotation
is seen in the central 50''.   The grey scale ranges from 1 to 50 mJy/beam whereas
the contours are plotted for 3,6,9,12,15 mJy/beam.
{\bf (b)} PV diagram along an axis parallel to the major axis passing through the northern HI loop,
the HI blob located off the 
eastern edge of NGC 2820, the streamer emerging
from NGC 2814 and NGC 2814.  The streamer and NGC 2814 although positionally
coincident, are kinematically distinct. The grey scale and contour levels
are the same as (a).}
\label{fig9}
\end{figure*}

A large one-sided HI loop is detected to the north of NGC 2820.  The loop extends out
to about 4.9 kpc along the rotation axis of the galaxy and has a lateral dimension
of about 17.5 kpc.  No counterpart is detected in the south of the galaxy. 
Moreover, we do not detect any radio continuum from the loop and
no H$\alpha$ emission is seen to be associated with
the loop.  In this section, we examine three possible scenarios for
the origin of the HI loop, namely a) starburst driven superwind  b) ram-pressure stripped
HI and c) tidally stripped HI.  We look for an origin which can
explain the observed constraints:  1) the absence of H$\alpha$ and radio continuum in the loop 
2) the one-sided nature of the loop  3) the symmetry of the loop 
about the rotation axis.  
Study of the velocity field of the HI loop gives the following inputs to
the above scenarios: 1) The HI loop 
is trailing the disk rotation.  The radial velocity
at the top of the loop is close to the systemic velocity of the galaxy (see Fig \ref{fig4} (b)). 2)
The line width increases along the loop and is largest at
the top of the loop (see Fig \ref{fig4} (c)).  3) The global HI profile 
and the HI profile of the loop are centred on the systemic velocity of the
galaxy and show no tail-like feature indicating that no gas is moving along the line of sight.

\subsection{Starburst driven superwind}
Gas is driven out of the disk by the underlying starburst to tens of kpc from the disk 
along the minor-axes and is known as superwind.  Superwind cones are commonly
observed in X-rays,  radio continuum and H$\alpha$ but infrequently in HI.  In case
of NGC 253, HI has been observed to be confined within the optical disk and
to outline the superwind cone of ionized gas in one half of the 
galaxy (Boomsma et al. \cite{boomsma}).  
Another study has shown that significant amounts of HI is
observed to be in the halos of spiral galaxies with active star formation
Fraternali et al. (\cite{fraternali}).  In case of NGC 2820, the HI halo is 
not observed to be significantly larger than the optical disk (see Fig \ref{fig7}(a)).
Anyway, here, we examine the possibility of the large HI loop
that we observe in NGC 2820 being a superwind and estimate its energetics. 

The mass of the HI loop is $5.5 \times 10^{8} M_\odot$ and 
the full width at zero intensity of the HI line is 160 km\,s$^{-1}$.
Assuming HI to be flowing along the surface of a superwind bi-cone
of half angle $45^{\circ}$, the deprojected outflow velocity is 113 km\,s$^{-1}$.
The kinetic energy contained in the outflowing HI is $7 \times 10^{55}$ ergs
and the dynamical age of the outflow is 34.7 million years.

We compare this with the energy contained in the supernovae in NGC 2820 to examine the
feasibility of this scenario.  
We follow Condon (\cite{condon1}) and use the non-thermal radio 
continuum emission from the central parts of NGC 2820 to estimate a supernovae rate 
of $\sim 0.007$ per year. 
This rate is fairly low compared to typical superwind galaxies 
e.g. 0.1 supernovae per year in NGC 1482, M82 and 0.1 to 0.3 in NGC 253.

If the kinetic energy imparted to the interstellar medium
by a single supernova explosion is $10^{51}$ ergs, then the kinetic energy available during the
dynamical age of the outflow is $3.4 \times 10^{56}$ ergs.  
However, note that the starburst would also drive other phases of the interstellar 
medium into the halo and hence only about
$1-10\%$ of the estimated kinetic energy will be seen in HI.  In this case, the energy available
in the supernovae is at best comparable to the kinetic energy of the HI loop.  
We arrive at a kinetic energy within a factor of few higher from the above if we include
the energy due to the stellar winds from massive young stars.  This was estimated from the
total FIR luminosity of NGC 2820 following the method 
by Heckman et al. (\cite{heckman}).  

Even if the energetics just about satisfy the starburst-driven origin of the loop, we note several
reasons below why we do not favour this scenario: 
(a) No H$\alpha$ emitting gas is found along the HI loop unlike superwind galaxies. 
(b) NGC 2820 is not classified as a starburst galaxy and does not show any obvious signatures
of being a starburst galaxy.
(c) No HI loop of such large dimensions appears to have been observed in any superwind galaxy.
(d) The HI loop is one-sided.  The starburst process which can drive such a large
loop in one direction should be energetic enough to drive it in the other direction also.

\subsection{Ram pressure stripping}

In this subsection, we examine the scenario of the observed HI loop as being the stripped HI
from NGC 2820 due to ram pressure (Gunn \& Gott \cite{gunn})  
exerted by the IGrM.  The reason we examine this possibility is the presence of
a few interesting features in the group members.
Firstly the southern edge of NGC 2820 shows a sharp cutoff in HI whereas 
the radio continuum extends beyond this cutoff (see Fig \ref{fig7} (b)).  
Secondly the northern part of NGC 2814 
shows a sharp cutoff both in radio continuum and HI whereas the optical disk does not
show any such effect.  Since ram pressure is expected to distort the various
disk components differently (Davis et al. \cite{davis}) and the sharp edges
can be caused by swept-back material due to ram pressure, we tend towards interpreting the above
features to be a signature of ram pressure of the IGrM. 

Thus it appears that ram pressure could have played an important 
role in the evolution of the group.
Hence, we examine the case of the HI loop seen to the north of NGC 2820 being a result of
ram pressure stripping.  We compare the
pressure due to the IGrM as the galaxy moves in it and the gravitational pressure of the disk of NGC 2820.
The relevant equation given by Gunn \& Gott (\cite{gunn}) is 

\begin{center}
$\rho\, v^2\, \ge\, 2\pi\, G\, \Sigma_*\, \Sigma_{gas}$\,\,\,\,  (\mbox{kg\,m$^{-1}$\,s$^{-2}$})
\end{center}

where $\rho$ is the IGrM mass density, $v$ is the velocity dispersion of the group,
G is the gravitational constant, $\Sigma_*$ is the surface mass density of stars,
$\Sigma_{gas}$ is the surface mass density of gas.
The line of sight velocity dispersion (Osmond \& Ponman \cite{osmond}) 
of this group is $162\pm73$ km\,s$^{-1}$. 
No X-ray emission has been detected from this group and the
upper limit on the bolometric X-ray luminosity is
$2.88 \times 10^{40}$ erg\,s$^{-1}$ (Mulchaey et al. \cite{mulchaey}).
If we assume that the temperature of the IGrM in this
late-type group is $2\times10^6$ K (Mulchaey et al. \cite{mulchaey3}) and the
medium is distributed over a sphere of radius 50 kpc,
then we arrive at upper limits on the electron
density of $8.8\times10^{-4}$ cm$^{-3}$ and mass of $1.15\times10^{10}$ M$_\odot$.
We assume a IGrM density of $4\times10^{-4}$ cm$^{-3}$ (well within the upper limit) and 
using v$^2=3*\sigma^2$ (Sarazin \cite{sarazin})  
we calculate a ram pressure of $\sim 6\times10^{-14}$ kg\,m$^{-1}$\,s$^{-2}$.  

We find an average stellar mass density of 133 $M_{\odot}$\,pc$^{-2}$ in the central
10 kpc region of NGC 2820 using the inclination-corrected total B-band magnitude
and the average $\gamma_B$ (light to mass ratio) factor 
given by Binney and Merrifield (\cite{binney}).  
For $N_H = 4.4 \times 10^{19}$ cm$^{-2}$, the surface mass density of HI is
0.32 $M_{\odot}$\,pc$^{-2}$.
These give the gravitational pressure of the disk to be
to be $\sim 8\times10^{-14}$ kgm$^{-1}$s$^{-2}$.  This is comparable to the
ram pressure acting on the system.  
If we estimate the gravitational pressure of the disk in the outer regions by
using the stellar mass density at the 25 mag-asec$^{-2}$ diametre (15.25 M$_\odot$\,pc$^{-2}$), 
which is close to where the loop is seen to emerge from the disk, then we find that the
ram pressure exceeds the gravitational pressure of the disk by a factor of few.
However this does not take into account the influence of dark matter in
the outer regions.

Mulchaey et al. (\cite{mulchaey2}, \cite{mulchaey}) have concluded from their X-ray observations of
many poor groups that the X-ray luminosity of the IGrM of groups with only late-type galaxies
is lower than if the group has at least one early-type member.  They infer 
that the late-type-only groups might have a lower temperature or a lower
density.  Our results seem to indicate that the IGrM should have densities that
are almost sufficient to morphologically influence NGC 2820.  Thus, the densities
should not be too low ($<< 4\times10^{-4}$) to be inconsequential.  

Assuming ram pressure has played a role in giving rise to the loop, we outline
two possible ways in which the loop could have formed.

\paragraph{Model 1:}
NGC 2820 has been classified as a barred galaxy (Bosma et al. \cite{bosma}).
Many barred galaxies show concentration of HI at the edge of the bar
and the loop like structure could be due to the neutral gas from this region
being stripped off in its interaction with the IGrM. 
In this scenario, the stripped HI lies well within the disk, just below the loop.  
The rotation of the disk causes a twist in the flow.
Since HI in this region is likely to be strongly bound to the disk, this model requires
extensive help from tidal effects in reducing the surface density of the neutral gas.
The kinematic features can be explained as: the trailing
velocity field could be due to a combination of vertical motion of gas stripped
from different regions in the disk combined with twisting due to galactic rotation.
The wider lines at the top of the loop could be because of gas acquiring higher random motions.
The central hole in the loop is expected since ram pressure cannot strip the high
density neutral gas in the central regions.

\paragraph{Model 2:}
The alternate scenario is that HI has been stripped from the edges of
the outermost spiral arms - one towards us and the other away from us.
Since NGC 2820 is an almost edge-on galaxy, this model is not distinguishable from the above.
This model can also explain the observed kinematic features of the loop.
The trailing velocity field will be due to the rotation velocity of the gas in the 
outer spiral arms which might not be the tangent points for those lines of sight.
The large line widths would again result from increased random motion as
the stripped gas meets the IGrM.  However given sufficient time, the central
hole in HI should get filled up in this model. 

With the present data, we cannot distinguish between the two models.  

We estimate
a velocity width of about 80 km\,s$^{-1}$ along the loop. 
This radial velocity is possibly dominated by rotation since NGC 2820 is a highly inclined
galaxy.  However, if we assume this to be the outflow velocity then for a loop
height of 4.9 kpc, this translates to an age of 60 million years.

Notice that the HI after reaching a certain height
seems to be falling back towards the disk (see Fig \ref{fig4} (a)).   
Vollmer et al. (\cite{vollmer}) in their simulations for investigating the role of
ram pressure stripping in the Virgo cluster have
found that ram pressure can cause a temporary increase in the central gas surface density
and in some cases even lead to a significant fraction of the stripped off atomic gas
to fall back onto the disk.  Such a process could be active in the case of NGC 2820. 

\subsection{Tidal stripping}
Interaction usually creates irregular morphology and tidal tails that
can stretch the spiral arms. 
The shape of the parent galaxy and the tidal tails can assume a 
variety of shapes after the interaction (for example, see simulations by
Toomre \& Toomre \cite{toomre}, Barnes \cite{barnes}, Howard et al. \cite{howard}).  
The members of Ho 124, especially 
the triplet, have undergone
close encounters which is evident from the numerous tidal features seen
in the system as described in section \ref{tidal}.  However it is not clear
how tidal interaction alone can give rise to the HI loop as observed in NGC 2820.
If one assumes that all the HI features seen to the north of the disk of NGC 2820 ie. the 
warped extension, the small protrusions, the loop and the streamer are parts of a tidal
tail, it is difficult to explain the origin of such a large distortion 
(stripped HI mass $>7\times10^8$ M$_\odot$) due to
a retrograde interaction with a galaxy (NGC 2814) whose HI content is less than that in the
HI loop.  Moreover no such tail is visible in the opposite direction which such a strong tidal
interaction should have produced.  Thus, it appears unlikely that the loop is a result
of tidal interaction alone.  
But on the other hand evolution of the system has obviously been affected by
tidal interactions e.g. Artamonov et al. \cite{artamonov} note
enhanced star formation due to the tidal interaction.   
Hence any model that explains the HI loop should 
include tidal interaction.  However it is  beyond the scope of this paper
to make any quantitative estimates of the tidal interaction which require
detailed simulations.

{\it We suggest that the loop could have been created by 
the combined effect of ram pressure and tidal forces acting on NGC 2820.}

\section{Discussion}
In this paper, we have described four main results arising from our observations, (i) the
steep radio continuum bridge between the triplet, (ii) the sharp cutoffs in different
galactic constituents observed in 
three members of the group,  (iii) one-sided HI loop in NGC 2820 and (iv) various signatures
left behind by the tidal interaction.   HI is detected from the bridge with
a mean column density of $4.4\times10^{19}$ cm$^{-2}$.
The bridge has a steep synchrotron spectrum with a spectral index of $-1.8^{+0.3}_{-0.2}$ 
and hence has large energy
losses caused by synchrotron and/or inverse Compton processes or a steep electron spectrum.  
From equipartition arguments, we find that relativistic particles and 
magnetic field dominate the bridge evolution.  
It contains a small fraction of the total sychrotron emission in the system and it
is interesting to contrast it with the Taffy galaxies in which the bridge emission
constitutes half of the total synchrotron emission in the system (Condon et al. \cite{condon}).

\begin{figure*}
\centering
\resizebox{\hsize}{!}{\includegraphics{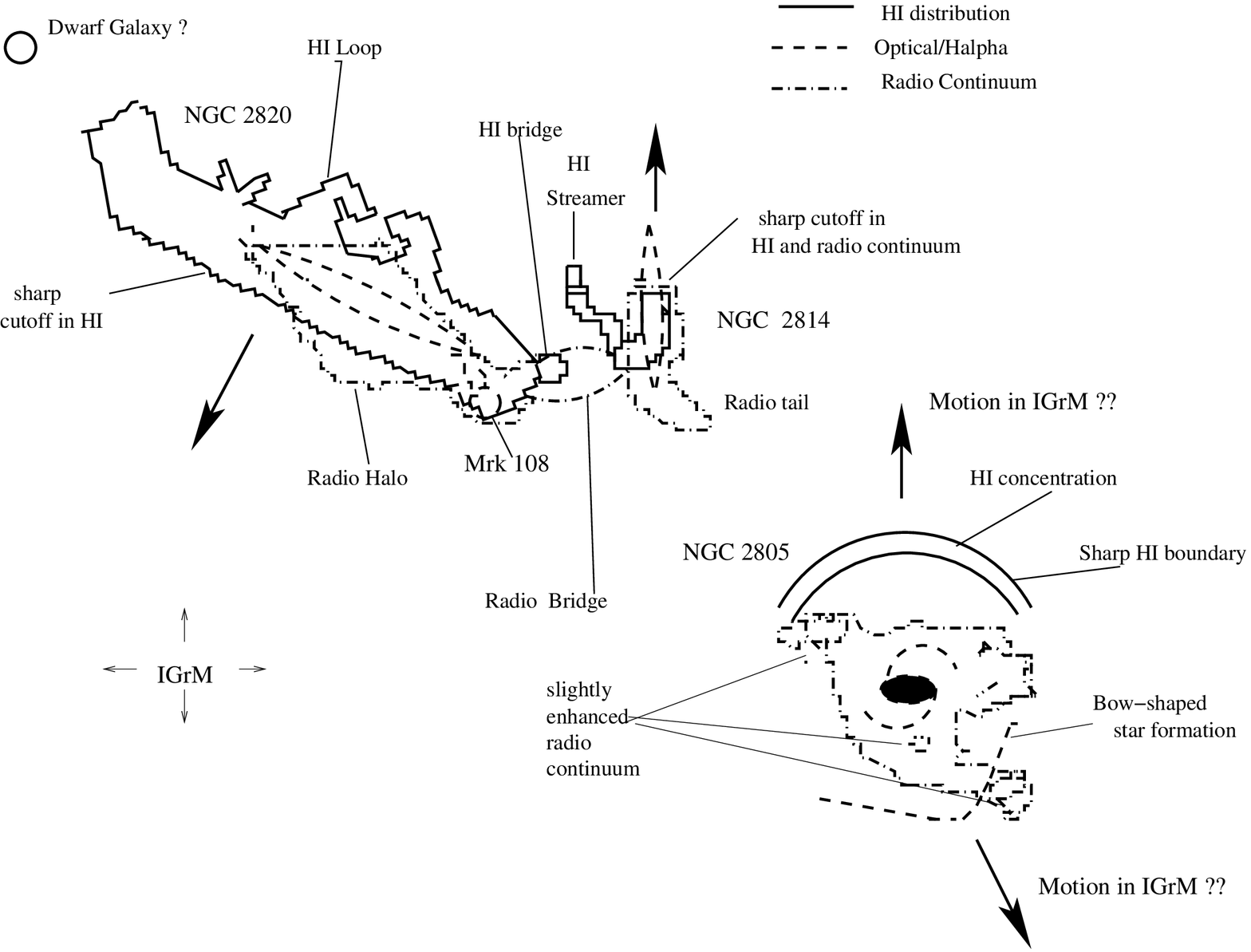}}
\caption{A Schematic of the group Holmberg 124.  The solid line represents the
observed HI 21cm distribution, the dashed line is for DSS optical distribution
and the dash-dotted line is for the radio continuum distribution in the group.
The thick arrows show the direction of motion of the galaxies in the group
as inferred from the sharp edge.  Note the ambiguity in the direction
of motion of NGC 2805.  Please note that the figure is not to scale. }
\label{fig10}
\end{figure*}

A sharp cutoff in HI (see Fig \ref{fig3} \& Fig \ref{fig5} (a)), 
radio continuum (see Fig \ref{fig1} (a)) 
or optical blue band (see Fig \ref{fig5}) is clearly 
evident in three members of Ho 124.  The fourth member (Mrk 108) is
small and too tidally disrupted.  The above is schematically summarized in Fig \ref{fig10}.
We suggest that the sharp boundaries are caused by motion of the
galaxies in the IGrM.
In the case of NGC 2814, the optical disk  appears
to be viewed edge-on with zero position angle
and the compression in radio continuum and HI is seen to be perpendicular 
to the major axis of the disk in the north in the sky plane.  Moreover
the radio continuum and HI are confined to well within the optical disk.
In NGC 2820, the HI in the southern side is sharply truncated.
The interaction between the triplet appears to have left behind a
trail of tidal debris like the streamer, the HI blobs and the
radio continuum tail of NGC 2814.  The HI loop could be a result of
both tidal effects and ram pressure.  If the tidal effects  have reduced the surface
density of HI in the disk of NGC 2820, then ram pressure of the IGrM should have been able to strip 
the outlying HI giving rise to the HI loop. 
The solid arrows near NGC 2820 and NGC 2814 in Fig \ref{fig10} 
indicate the direction of motion of those galaxies as we can deduce in the sky plane
from the sharp truncations.  
However, the ambiguity of the sharp cutoff in HI in the north and the enhanced
star formation in the south of NGC 2805 makes its direction of motion in the IGrM
ambiguous.  If we interpret the star formation ridge to be due to the interaction
with IGrM, then NGC 2805 appears to be moving towards the south-west.  If this is true
then the three galaxies seem to be moving in different directions.  However if
we interpret the sharp HI boundary in the north of NGC 2805 as due to ram pressure,
then the galaxy is moving towards the north.  In short, we cannot comment on the direction
of motion of NGC 2805 from the existing observations.

We compared the ram pressure and the equivalent pressure in the different phases
of the interstellar medium in NGC 2820 to examine its effect on different constituents
of the medium which show distinct morphologies and hence understand the observational picture sketched in
Fig \ref{fig10}.  Assuming a typical particle
density of $4\times10^{-4}$ cm$^{-3}$ for the IGrM, we estimated 
a ram pressure of 4170 cm$^{-3}$\,K for the IGrM. 
We estimate a magnetic pressure $\sim 5000$ cm$^{-3}$\,K in the radio continuum halo.  
The two pressures are comparable.
For the H$\alpha$ seen in the extended-diffuse ionized gas (e-DIG), 
Miller \& Veilleux (\cite{miller}) estimate an emission measure of about 8 pccm$^{-6}$ 
and a size of 2 kpc.  
If we assume a filling factor of 0.1 (see Fig 10 in Miller \& Veilleux (\cite{miller}))
for e-DIG then for a temperature of $10^4$ K, the pressure 
in the gas is 2000 cm$^{-3}$\,K and the difference by a factor of two could easily
be due to an incorrect filling factor assumed by us.  The pressure in the e-DIG 
is also comparable to the ram pressure. 
Lastly we estimate the pressure in the HI gas. 
Since the sharp edge in HI is unresolved by
our beam, we used a column density of $3.2\times10^{20}$ cm$^{-2}$ (the
second contour) to estimate an atomic density of $0.02$ cm$^{-3}$ for
a volume filling factor of 0.4 (taken from our Galaxy for warm neutral medium). 
If we use a temperature of 5000 K which is  typical of the warm neutral medium in our Galaxy,
then the pressure is 100 cm$^{-3}$\,K. 
The HI pressure is significantly lower than the ram pressure due to the IGrM.  
Thus we can understand the observed picture where HI has a swept-back appearance
in the south of NGC 2820 which could be due to ram pressure affecting 
whereas the radio continuum and H$\alpha$ seem to extend out beyond HI and
are either expanding or in pressure equilibrium with the IGrM.  For this the
galaxy should be in motion along the south-west as shown by the solid arrow
in Fig \ref{fig10}.

The HI distribution in NGC 2805 is asymmetric with enhanced column densities 
in clumps towards the north (see Fig \ref{fig5} (a)).
The HI is extended east-west along the highly-disturbed northern optical spiral arm 
and is confined to the optical disk.  The radio continuum emission
in NGC 2805 is fairly weak (see Fig \ref{fig5} (b)), with localised peaks in the south
and northeast, and bearing little resemblance either to the optical or to the HI
distribution.  
The central region of the galaxy 
is bright in the JHK near-IR bands and in the optical band but faint in
radio continuum (Fig \ref{fig5} (b)) which is intriguing.
In short, this massive galaxy appears to be highly disturbed, the reason for
which is not clear from existing observations.  

Thus, we end with a picture (somewhat speculative) of this group 
as derived from the present and other existing observations as shown in Fig \ref{fig10}.
NGC 2820 has probably undergone a
retrograde tidal encounter with NGC 2814 which has left behind a trail of tidal debris.  A 
HI streamer probably detached from NGC 2820 is seen projected onto NGC 2814 (see Fig \ref{fig10})
but is kinematically distinct from NGC 2814. 
The interaction has also probably given rise to a bridge connecting the two galaxies and
a tail of radio continuum emission in the south of NGC 2814.  
Star formation has been triggered in south-west parts
of the disk of NGC 2820, in Mrk 108 and in the southern parts of NGC 2814
(see Fig \ref{fig7} (c)).  
Using about half the upper limit on the electron density estimated from the
upper limit on the X-ray emission (Mulchaey et al. \cite{mulchaey}), 
we estimate the ram pressure force of
the IGrM to be comparable to the gravitational pull of the disk of NGC 2820.
Since tidal interaction has obviously
influenced the group, we suggest that the loop could have formed by ram pressure
stripping if tidal effects had reduced the surface density of HI in NGC 2820.
We suggest that the HI loop which is several kpc high and across could have been produced by
the combined effects of ram pressure and tidal forces.  Moreover we find sharp truncations
to the HI in some of the group members which we believe supports the ram pressure explanation.
If we assume that this is true then NGC 2814 is probably moving towards the north
and NGC 2820 towards moving in the south-east direction
(solid arrows in Fig \ref{fig10}).   From the existing observations there is ambiguity in
the direction of motion of NGC 2805.  
Considering all this, we suggest that the group evolution is being influenced
by both tidal forces due to the mutual interactions and ram pressure due to the
motion of the galaxies in an IGrM.

Since the IGrM is not detected in X-rays (Mulchaey et al. \cite{mulchaey}) 
but we tend to believe shows detectable effect on the galaxies in Ho 124, 
we suggest that the IGrM densities in this group should not be too low. 
Mulchaey et al. (\cite{mulchaey2},\cite{mulchaey}) 
have suggested that the non-detection of X-rays in late-type groups could
be due to lower temperatures or densities.  
The detection of a 0.2 keV (Wang \& McGray \cite{wang}) 
IGrM in the Local group which is a late-type group indicates that such groups do have a
IGrM.  It is more difficult to detect lower temperature gas in groups
for several reasons like the enhanced
absorption of such soft X-rays by galactic HI and the increased strength of 
the X-ray background.  One clearly needs to explore other avenues of detecting
this gas. 

\section{Summary}

\begin{itemize}

\item We detect the faint radio continuum bridge at
330 MHz connecting NGC 2820+Mrk 108 with NGC 2814 which was
first detected by van der Hulst \& Hummel (\cite{hulst}) at 1465 MHz.  The bridge has a 
a spectral index of $-1.8^{+0.3}_{-0.2}$ which is steeper than
the $-0.8$ quoted by van der Hulst \& Hummel (\cite{hulst}).
HI is detected from most of the bridge
at a velocity close to the systemic velocity of NGC 2820 and has a mean column density of
$4.4 \times 10^{19}$ cm$^{-2}$.  No H$\alpha$ emission is associated with the bridge. 

\item We detect radio contiuum from all the members of the group.  
A radio halo is clearly detected around NGC 2820 in the radio
continuum with a $10\%$ peak flux density extent of 4.2 kpc at 330 MHz and 
a spectral index of $-1.5$.  
A radio halo is also detected around NGC 2814.  The radio continuum at
330 MHz from NGC 2805 is fairly weak bearing little resemblance to either the HI
distribution or the optical emission.  The centre of the galaxy is intriguingly faint.

\item HI is detected from all the galaxies in the group. 
The heliocentric systemic velocity of NGC 2820 is 1577 km\,s$^{-1}$ and its
rotation velocity is 175 km\,s$^{-1}$.  
The linear extent of the HI disk of NGC 2820 is about 48 kpc and its 
HI mass is $6.6 \times 10^{9}$  M$_\odot$.
The HI emission associated with Mrk 108 is clearly detected at 1417 km\,s$^{-1}$
and it encloses a HI mass of $6.1\times10^7$ M$_\odot$.

\item We detect various tidal features close to NGC 2814.
The radio continuum disk and HI disk of NGC 2814 are tilted with respect
to the optical disk.
A HI streamer is seen to emerge from the south of NGC 2814 but the two
are
are kinematically distinct.  The velocity field of the streamer 
is similar to parts of NGC 2820 close to it.
The streamer has a sky plane extent of 12.6 kpc and encompasses an HI mass of
$1.3\times10^8$ M$_\odot$.
A tail emerging from the south of NGC 2814 and extending westwards
is detected in the radio continuum.  The tail has a spectral index of $-1.6$.

\item We detect HI gas located about 11.5 kpc to the north-east of NGC 2820 whose
dynamical mass is $1.4\times10^9$ M$_\odot$ and which might possibly be a tidal dwarf galaxy.
However, deep H$\alpha$ observations are required to confirm this.
The velocity of this gas is similar to the velocity field of the part of NGC 2820
closest to it.

\item We observe a sharp cutoff in HI on the southern rim of NGC 2820 and a
sharp truncation in HI and radio continuum to the north of NGC 2814.
We suggest that these features could be a result of ram pressure due to motion of 
the galaxies in the IGrM along the solid arrows shown in Fig \ref{fig10}  
since simple estimates of pressure in different components of the interstellar medium
in NGC 2820 suggest that ram pressure exceeds the pressure in HI by a
factor of many.  However this needs to be verified.

\item We report detection of a gigantic HI loop arising to the north of NGC 2820.  
The loop is $\sim 17.5$ kpc across and rises up to $\sim 4.9$ kpc.  It
encompasses an HI mass of $6\times10^8$ M$_\odot$.  
No radio continuum or H$\alpha$ emission is associated with this loop.   
We present possible origin scenarios which include a central starburst,
ram pressure stripping and tidal stripping.  We do not favour a central
starburst mainly because of the absence of detectable ionized
gas in the loop.  We tend to favour the ram pressure scenario.
Using the upper limit on the X-ray luminosity from Ho 124 (Mulchaey et al. \cite{mulchaey}),
we estimate an upper limit on the electron density of $8.8\times10^{-4}$ cm$^{-3}$.
Our calculations using half this electron density
show that ram pressure force of the IGrM is comparable to the
gravitational pull of the disk.  Hence we 
suggest that this loop could have been formed due to ram pressure stripping if 
tidal forces had reduced the surface density of HI in NGC 2820.

\item The group under study exhibits multiple signatures of tidal interaction and
possibly ram pressure.  Thus, we suggest that the evolution 
of Ho 124 may be governed by both tidal
interaction and ram pressure due to the motion of the galaxies in the IGrM.

\end{itemize}

\begin{acknowledgements}
We thank the staff of the GMRT that made these observations possible. 
GMRT is run by the National Centre for Radio Astrophysics of the Tata Institute of Fundamental Research.
This research has made use of the NASA/IPAC Extragalactic Database (NED) which is 
operated by the Jet Propulsion Laboratory, California Institute of Technology, 
under contract with the National Aeronautics and Space Administration.
The Digitized Sky Survey was produced at the Space Telescope Science Institute 
under U.S. Government grant NAG W-2166. The images of these surveys are based 
on photographic data obtained using the Oschin Schmidt Telescope on Palomar Mountain 
and the UK Schmidt Telescope. The plates were processed into the present compressed 
digital form with the permission of these institutions.
NGK acknowledges a discussion with Amitesh Omar.  We thank D. J. Saikia for going through
the manuscript and providing useful inputs.  We thank the anonymous referee for 
comments which have helped improve the paper.  

\end{acknowledgements}

\end{document}